\documentclass[seceq]{ptptex}
\usepackage{graphicx,bm,wrapft}
\def\fsl#1{\setbox0=\hbox{$#1$}                 
   \dimen0=\wd0                                 
   \setbox1=\hbox{/} \dimen1=\wd1               
   \ifdim\dimen0>\dimen1                        
      \rlap{\hbox to \dimen0{\hfil/\hfil}}      
      #1                                        
   \else                                        
      \rlap{\hbox to \dimen1{\hfil$#1$\hfil}}   
      /                                         
   \fi}                                         %

\newcommand{\Tr}{\mbox{Tr}}

\newcommand{\diag}{\mbox{diag}}
\newcommand{\VEV}[1]{\langle #1 \rangle}

\notypesetlogo

\title{Gluonic phases and phase diagram in neutral two flavor dense QCD}

\author{Michio \textsc{Hashimoto}$^{1,}$\footnote{
E-mail: michioh@eken.phys.nagoya-u.ac.jp}
and 
V.~A. \textsc{Miransky}$^{2,}$\footnote{E-mail: vmiransk@uwo.ca; 
On leave from Bogolyubov Institute for Theoretical Physics, 03143, 
Kiev, Ukraine. }
}

\inst{$^1$Department of Physics, Nagoya University, Nagoya, 464-8602, Japan \\
$^2$Department of Applied Mathematics, University of Western Ontario, \\
London, Ontario N6A 5B7, Canada
}

\abst{
A numerical analysis of several phases in the cold
neutral two flavor dense quark matter is realized. In the analysis, 
besides the normal, single plane wave LOFF, and color superconducting
2SC/g2SC phases, we also include two gluonic phases (the gluonic cylindrical
phase II and the gluonic color-spin locked one). 
It is shown that these two gluonic phases are 
dynamically realized and, on the basis of this analysis,
the phase diagram with respect to the coupling strength
in this medium is suggested. 
In particular, it is shown that
the gluonic phases are energetically favorable 
in a wide region of the parameter space and they can resolve
the problem of the 2SC/g2SC chromomagnetic instability
in the whole region where it exists.
On the other hand, there also exists
a window where the LOFF phase is stable. The 2SC state and the 
normal one are realized in the strong and weak coupling 
regimes, respectively, as was expected.
}

\begin{document}
\maketitle

\section{Introduction}

In this paper, we will present a self-consistent numerical analysis of
several phases in 
the cold neutral two flavor dense quark matter. In particular,
we analyze gluonic phases introduced in 
Ref.~\citen{Gorbar:2005rx} and described recently in detail 
in Ref.~\citen{Gorbar:2007vx}. 
On the basis
of this analysis, it is shown that a) certain gluonic phases 
connected with a first order phase transition 
are dynamically realized and
b) the phase diagram with respect to the coupling strength
in this medium is suggested.
In particular, it is shown that
the gluonic phases are energetically favorable
in a wide region of the parameter space and they can resolve
the problem of the 2SC/g2SC chromomagnetic instability
in the whole region where it exists.

It is expected that at sufficiently high baryon density, cold quark
matter should be in a color superconducting state (for reviews,
see Ref.~\citen{review}).
On the other hand, it was suggested long ago that quark matter might
exist inside the central region of compact stars.\cite{quark_star}
This is one of the main reasons why the dynamics of
the color superconductivity has been intensively studied.

Bulk matter in compact stars should be in $\beta$-equilibrium,
providing by weak interactions,
and be electrically and color neutral.
The electric and color neutrality conditions play a crucial role
in the dynamics of quark pairing.\cite{Iida:2000ha,Alford:2002kj,Steiner:2002gx,Huang:2002zd,Huang:2003xd,Abuki:2004zk,Ruster:2005jc,Blaschke:2005uj}
Also, in the dense quark matter,
the strange quark mass cannot be neglected.
These factors lead
to a mismatch $\delta\mu$ between the Fermi momenta of
the pairing quarks. 

The mismatch $\delta\mu$ tends to destroy the gapped (2SC) and 
gapless (g2SC) two-flavor color superconducting phases and the
color-flavor locked (CFL) and gapless color-flavor locked 
(gCFL) three-flavor phases. 
Establishing the genuine
ground state (vacuum) for realistic values of $\delta\mu$ in dense QCD
is one of the central problems in the field. 
Besides the gluonic phases,\cite{Gorbar:2005rx,Gorbar:2007vx} \
a number of other candidates for the ground state have been proposed.\cite{Alford:2000ze,Bowers:2002xr,Reddy:2004my,Huang:2005pv,Kryjevski:2005qq,Zhang:2006rp,Gatto:2007ja}

An important factor in solving this problem 
is removing instabilities observed in gluon (color plasmon)
channels in the 2SC/g2SC and gCFL 
phases.\cite{Huang:2004bg,Casalbuoni:2004tb,Gorbar:2006up}
As was revealed in Ref.~\citen{Huang:2004bg},
the 2SC/g2SC phases 
suffer from a chromomagnetic
instability connected with the presence of imaginary
Meissner masses of gluons.
While the 8th gluon has an imaginary Meissner mass only
in the g2SC phase, with the diquark gap
$\bar{\Delta} < \delta\mu$ (an intermediate coupling regime),
the chromomagnetic instability for the 4-7th gluons appears also in
a strong coupling regime in the 2SC phase, with
$\delta\mu < \bar{\Delta} < \sqrt{2}\delta\mu$.
Later a chromomagnetic instability was also found in
the three-flavor gCFL phase.\cite{Casalbuoni:2004tb}

Meissner and Debye masses are screening (and not pole) ones.
It has been recently revealed in Ref.~\citen{Gorbar:2006up}
that the chromomagnetic instabilities
in the 4-7th and 8th gluonic channels in the 2SC/g2SC phases
correspond to two very
different tachyonic spectra of plasmons.
It is noticeable that while
(unlike the Meissner mass)
the (screening) Debye mass for an electric mode remains real for
all values of $\delta\mu$ both in the 2SC and g2SC 
phases,\cite{Huang:2004bg} \
the tachyonic plasmons occur both for the magnetic and electric 
modes.\cite{Gorbar:2006up} The
latter is important since it clearly shows that this instability is connected
with vector-like excitations: Recall that
two magnetic modes correspond to two
transverse components of a plasmon, and one electric mode
corresponds to its longitudinal component.
This form of the plasmon spectrum leads to the unequivocal conclusion
about the existence of vector condensates of gluons
in the ground state
of two flavor quark matter with
$\bar{\Delta} < \sqrt{2}\delta\mu$, thus
supporting the scenario with gluon condensates (gluonic phases)
proposed in Refs.~\citen{Gorbar:2005rx,Gorbar:2007vx}.

Besides the instabilities in gluon channels, 
there exists an instability 
reflected in a negative velocity squared $v^2< 0$
of the physical
diquark excitation (diquark Higgs mode) in the g2SC phase
\cite{Hashimoto:2006mn} (a similar instability was
discussed in Refs.~\citen{Iida:2006df,Giannakis:2006gg}).  
This problem should be also resolved in the
genuine ground state. 

In Refs.~\citen{Gorbar:2005rx,Gorbar:2007vx},
by using the Ginzburg-Landau (GL) approach in the hard dense loop (HDL)
approximation,
it was demonstrated the existence of a particular gluonic phase in 
the vicinity of
the critical point 
$\delta\mu \simeq \Delta/\sqrt{2}$.
Since the GL approach works only for second order phase
transitions, a specific ansatz, leading to such a transition, was
considered. It is:
\begin{equation}
  \mu_8 \equiv \frac{\sqrt{3}}{2} g\VEV{A_0^{8}}, \quad
  \mu_3 \equiv g\VEV{A_0^{3}}, \quad 
  B \equiv g \VEV{A_z^{6}}, \quad
  C \equiv g \VEV{A_z^{1}},
\label{cyl_glu_1}
\end{equation}
where $A_\mu^{a}$'s are gluon fields and
$g$ denotes the QCD coupling constant. In this phase,
the symmetry $SU(2)_c \times \tilde{U}(1)_{em} \times SO(3)_{\rm rot}$
is spontaneously broken down to $SO(2)_{\rm rot}$. Here $SU(2)_c$ and
$\tilde{U}(1)_{em}$ are the color and electromagnetic symmetries 
in the 2SC state, and $SO(3)_{\rm rot}$ is the rotational group
(recall that in the 2SC state the color $SU(3)_c$ is broken down
to $SU(2)_c$). In other words, the solution (\ref{cyl_glu_1})
describes an anisotropic medium in which the color and electric
superconductivities coexist. It is noticeable that there are
chromoelectric field strength condensates $\VEV{E^2_z}$ and $\VEV{E^7_z}$
in this phase, reflecting
its non-Abelian nature.
Since solution (\ref{cyl_glu_1})
is cylindrically symmetric, it was called the gluonic cylindrical phase I
in \citen{Gorbar:2007vx} (for the gluonic cylindrical phase II, 
see below).

The first important step in realizing a numerical analysis in the
gluonic phase outside the scaling region around 
$\delta\mu \simeq \bar{\Delta}/\sqrt{2}$ was done in 
Refs.~\citen{Fukushima:2006su,Kiriyama:2006ui}. 
In that analysis, some approximations were made:
the color chemical potentials $\mu_3$ and $\mu_8$
were ignored and it was
assumed that the mismatch
$\delta\mu$ and the diquark gap $\bar\Delta$
in the gluonic phase are the same as those
in the 2SC/g2SC one. 
However, it is unclear whether these approximations
are justified in the gluonic phase.
For example, while  in the 2SC/g2SC phases $\mu_3 = 0$ and $\mu_8$ is 
small,\cite{Gerhold:2003js} \
it is questionable whether they
can be neglected in the gluonic one. As to $\delta\mu$ and 
$\bar\Delta$ in the gluonic phase, nothing is known about them
away from the critical point either.

The main goal of the present paper is a numerical self-consistent analysis
of the gluonic phases. Unfortunately, at present it is very hard
to realize such a study in the phase (\ref{cyl_glu_1}). The point
is that
as was shown in Ref.~\citen{Gorbar:2007vx}, outside the scaling region, 
two additional gluon condensates, 
$\mu_4 \; (=g\VEV{A_0^4})$ and $\mu_5 \; (=g\VEV{A_0^5})$,
cannot be ignored.
With the four condensates in Eq. (\ref{cyl_glu_1}) and these two ones, 
plus two equations for $\delta\mu$ and $\bar{\Delta}$,
the numerical analysis in this phase is very involved. 
Fortunately, as was pointed out in \citen{Gorbar:2007vx} (based on the 
description of possible symmetry breaking structures in dense QCD), there may
exist other gluonic phases (presumably connected with first order
phase transitions). In some of them, it is easier to
realize a numerical analysis than in the gluonic cylindrical phase I.

Here we will be interested in the following two ones. 
First, a phase with $C = 0$ in Eq. (\ref{cyl_glu_1}) is not
excluded:
\begin{equation}
  \mu_8 = \frac{\sqrt{3}}{2} g\VEV{A_0^{8}}, \quad
  \mu_3 = g\VEV{A_0^{3}}, \quad
  B = g \VEV{A_z^{6}} \, .
 \label{cyl_glu_2}
\end{equation}
It was called the gluonic cylindrical phase II.\cite{Gorbar:2007vx} 
In this phase, while the rotational $SO(3)$
is also spontaneously broken down to $SO(2)_{\rm rot}$, the electromagnetic
$U(1)$ is preserved. We emphasize that since a chromoelectric field 
strength condensate $\VEV{E^7_z}$ is generated in this phase, it also possesses
a non-Abelian nature.

The second phase that will be analyzed here is the so called 
gluonic color-spin locked (GCSL) phase with the condensates 
\begin{equation}
  \mu_8 = \frac{\sqrt{3}}{2} g\VEV{A_0^{8}}, \quad
  K \equiv g \VEV{A_y^{4}} = g \VEV{A_z^{6}}.
 \label{csl_glu}
\end{equation}
In this phase, the symmetry
$SU(2)_c \times \tilde{U}(1)_{em} \times SO(3)_{\rm rot}$
is spontaneously broken down to the $SO(2)_{\rm diag}$ one,
with the generator which is an appropriate 
linear combination of $\sigma^2$ in the color $SU(2)_c$ and
the generator of spatial $y-z$ rotations. It is noticeable
that there exist both chromoelectric and chromomagnetic
field strengths condensates in the GCSL phase. Also, since
the electromagnetic $U(1)$ is spontaneously broken in this phase,
it (like the gluonic cylindrical phase I) describes an
anisotropic superconductor. 

It is quite nontrivial whether or not the gluonic phases 
(\ref{cyl_glu_2}) and (\ref{csl_glu}) connected with the first
order phase transition are
{\it dynamically} realized.
Fortunately, there are not so many parameters in these phases
that will allow us 
to perform numerical calculations for solving 
self-consistently both the neutrality 
conditions for $\mu_3$, $\mu_8$, $\delta\mu$ and 
the gap equations for $\Delta$, $B$ and/or $K$ (the five coupled
equations in the gluonic cylindrical phase II and the four
coupled equations in the GCSL phase). 

As a benchmark, we will also study the single plane LOFF 
phase,\cite{Alford:2000ze,Giannakis:2004pf,Gorbar:2005tx,He:2006vr} \
which is gauge equivalent to~\cite{Gorbar:2005rx,Gorbar:2005tx}
\begin{equation}
  \mu_8 = \frac{\sqrt{3}}{2} g\VEV{A_0^{8}}, \quad
 q \equiv \frac{1}{2 \sqrt{3}}\, g \VEV{A_z^{8}} \, . 
 \label{single_loff} 
\end{equation}
We will determine which of the phases 2SC/g2SC and
(\ref{cyl_glu_2}), (\ref{csl_glu}), and (\ref{single_loff}) is
energetically favorable in the bulk of the parameter space.

It turns out that there indeed exist dynamical solutions corresponding
to the gluonic cylindrical phase II and the GCSL one.
Actually, these phases are energetically 
more favorable than the single plane wave LOFF and 2SC/g2SC phases
in a wide region of the parameter space.
For realistic values $\mu=400$ MeV and $\Lambda=635.5$ MeV, 
where $\mu$ and $\Lambda$ are the quark chemical potential and the cutoff  
in the (gauged) Nambu-Jona-Lasinio (NJL) model, 
the gluonic cylindrical II and GCSL phases are more stable
than the normal, LOFF and g2SC/2SC phases in the wide interval 
$6.7 \times 10^1 \mbox{ MeV} < \bar{\Delta}_0 < 1.6 \times 10^2 \mbox{ MeV}$.
Here we introduced the 2SC gap parameter 
$\bar{\Delta}_0$ defined at $\delta\mu=0$
(it essentially corresponds to the strength of
the four-diquark-coupling constant in the (gauged) NJL model). 

On the other hand, the neutral normal phase with no condensates
is the ground state at $\bar{\Delta}_0 < 6.5 \times 10^1 \mbox{ MeV}$
(weak coupling regime) while 
the single plane LOFF phase is favorable in the narrow window 
$6.5 \times 10^1 \mbox{ MeV} < \bar{\Delta}_0 < 6.7 \times 10^1 \mbox{MeV}$. 
At $\bar{\Delta}_0 > 1.6 \times 10^2 \mbox{ MeV}$ 
(strong coupling regime),
the 2SC phase is realized. 

A noticeable point is that 
the gapless g2SC solution exists only at 
$92.2 \mbox{ MeV} < \bar{\Delta}_0 < 134.6 \mbox{ MeV}$, i.e.,
it occurs later (at larger values of $\bar{\Delta}_0$) than the
gluonic phases. This strongly
suggests that the gluonic phases can resolve the problem of the
chromomagnetic and plasmon instabilities in the 2SC/g2SC phases.

\section{Model}

As in Refs.~\citen{Gorbar:2005rx,Gorbar:2007vx},
we will use the gauged Nambu-Jona-Lasinio (NJL) model with 
two light (up and down) quarks, whose current masses will be neglected.
The Lagrangian density is given by
\begin{equation}
  {\cal L} = \bar{\psi}(i\fsl{D}+\bm{\mu}_0\gamma^0)\psi
  +G_\Delta \bigg[\,(\bar{\psi}^C i\varepsilon\epsilon^\alpha\gamma_5 \psi)
           (\bar{\psi} i\varepsilon\epsilon^\alpha\gamma_5 \psi^C)\,\bigg]
  -\frac{1}{4}F_{\mu\nu}^{a} F^{a\,\mu\nu} ,
 \label{Lag}
\end{equation}
where
\begin{equation}
  D_\mu \equiv \partial_\mu - i g A_\mu^{a} T^{a}, \;
  F_{\mu\nu}^{a} 
  \equiv \partial_\mu A_\nu^{a} - \partial_\nu A_\mu^{a} +
  g f^{abc} A_\mu^{b} A_\nu^{c}
\end{equation}
with $\varepsilon \equiv \varepsilon^{ij}$ 
and $\epsilon^{\alpha} \equiv \epsilon^{\alpha\beta\gamma}$ are the totally 
antisymmetric tensors in the flavor and color spaces, respectively;
$i,j = u, d$ and $\alpha, \beta, \gamma = r$ (red), $g$ (green), $b$ (blue). 
The matrices $T^{a}$'s are the generators of $SU(3)$ 
in the fundamental representation,
and $f^{abc}$'s are the structure constants of $SU(3)$.
The charge-conjugate spinor $\psi^C$ is 
$\psi^C \equiv C \bar{\psi}^T$ with $C = i\gamma^2\gamma^0$.
We do not introduce photon field and therefore the electric charge
is connected with a global $U(1)$ symmetry in the model.
On the other hand, the whole theory contains free electrons,
although we do not show them explicitly in Eq.~(\ref{Lag}).
In $\beta$-equilibrium, the chemical potential matrix $\bm{\mu}_0$, 
connected with the conserved global quark and electric charges, 
is 
\begin{equation}
  \bm{\mu}_0 = \mu - \mu_e Q_{\rm em} 
  = \bar{\mu}_0 - \delta\mu \tau_3.
 \label{mu}
\end{equation}
Here $\mu$ and $\mu_e$ are 
the quark and electron chemical potentials, 
$\bar{\mu}_0 \equiv \mu - \frac{\delta\mu}{3}$ and
$\delta \mu \equiv \frac{\mu_e}{2}$
(the baryon chemical potential $\mu_B$ is $\mu_B \equiv 3\mu$.)
The matrix $\tau_3 =\diag(1,-1)$ acts in the flavor space.
Also, in the 2SC/g2SC phases,
the color neutrality condition
requires a non-zero color chemical potential 
$\mu_8$.\cite{Gerhold:2003js}

The following comment is in order. In the full theory, the four-quark 
coupling constant $G_\Delta$ (or $\bar{\Delta}_0$) is not a free
parameter but expressed through the QCD coupling $g$ and the quark chemical
potential $\mu$. Needless to say, that at
present this relation for $\bar{\Delta}_0$ is unknown. Because of
that, we will treat $\bar{\Delta}_0$ as a free parameter, allowing
it to vary in a wide interval that includes its typical values
considered in the literature.

Let us introduce the diquark field
$\Phi^\alpha \sim i\bar{\psi}^C\varepsilon \epsilon^\alpha \gamma_5 \psi$.
We will use a gauge in which  
\begin{equation}
  \Phi^r=0, \quad  \Phi^g=0, \quad \Phi^b=\Delta ,
  \label{2SC}
\end{equation}
where the anti-blue component $\Delta$ is real. The gap $\bar{\Delta}$
is given by the vacuum expectation value (VEV) of $\Delta$,
$\bar{\Delta} \equiv \VEV{\Delta}$.

In the presence of both diquark field $\Delta$ and
gluon fields $A_\mu^a$,
by using the Nambu-Gor'kov spinor
\begin{equation}
  \Psi \equiv \left(\begin{array}{@{}c@{}} \psi \\ \psi^C \end{array}\right),
\end{equation}
the inverse fermion propagator $S_g^{-1}$ can be written as
\begin{equation}
  S_g^{-1}(P) = \left(
  \begin{array}{cc}
  [G_{0,g}^+]^{-1} & \Delta^- \\ \Delta^+ &  [G_{0,g}^-]^{-1} 
  \end{array}
  \right) ,
  \label{Sg-inv}
\end{equation}
with
\begin{eqnarray}
&&  [G_{0,g}^+]^{-1}(P) \equiv
  (p_0+\bm{\mu}_0)\gamma^0
  -\vec \gamma \cdot \vec p + g \fsl{A}^a T^a, \\[2mm]
&&  [G_{0,g}^-]^{-1}(P) \equiv
  (p_0-\bm{\mu}_0)\gamma^0
  -\vec \gamma \cdot \vec p - g \fsl{A}^a (T^a)^T ,
\end{eqnarray}
\begin{equation}
  \Delta^- \equiv -i\varepsilon\epsilon^b\gamma_5\Delta, \qquad
  \Delta^+ \equiv \gamma^0 (\Delta^{-})^\dagger \gamma^0 =
  -i\varepsilon\epsilon^b\gamma_5\Delta,
\end{equation}
where $P^\mu \equiv (p_0,\vec p)$ is the energy-momentum four vector.
Note that in this case the color neutrality requires in principle 
VEVs of the time components of all gluon fields: 
\begin{equation}
 \mu_{1,2,\cdots,7} \equiv g\VEV{A_0^{1,2,\cdots,7}} , \quad
 \mu_8 \equiv \frac{\sqrt{3}}{2}\, g\VEV{A_0^{8}} \, .
\end{equation}

We will utilize the fermion one-loop approximation, closely related
to the HDL one, in which only
the dominant one-loop quark contribution is taken into account
while the contribution of gluon loops is neglected. On the other hand,
we keep the tree contribution of gluons. This is because we want to 
check how sensitive are the results to the choice of the value of the
QCD coupling $g$.

In this approximation, 
the effective potential including both gluon and diquark fields is:
\begin{equation}
  V_{\rm eff} =
   \frac{\Delta^2}{4G_\Delta} 
  +\frac{g^2}{4}f^{a_1a_2a_3}f^{a_1a_4a_5}
    A_\mu^{a_2} A_\nu^{a_3} A^{a_4\,\mu} A^{a_5\,\nu}
  - \frac{\mu_e^4}{12\pi^2}
  - \frac{1}{2}\int\frac{d^4 P}{i(2\pi)^4}\Tr\ln S_g^{-1}(P) ,
  \label{V}
\end{equation}
where we also added the free electron contribution.
The following subtraction scheme will be used:
\begin{eqnarray}
  V_{\rm eff}^R(\mu,\delta\mu,\mu_a;\Delta,\vec A^{\,a}) &=& 
  V_{\rm eff}(\mu,\delta\mu,\mu_a;\Delta,\vec A^{\,a})
 \nonumber \\ && \hspace*{-1cm}
+ \frac{1}{2}\int\frac{d^4 P}{i(2\pi)^4}\Tr\ln 
   S_g^{-1}(P; \mu=\delta\mu=\mu_a=\Delta=0,\vec A^{\,a}),
  \label{V_R}
\end{eqnarray}
in which a spurious mass term $\Lambda^2 |\vec A^{\,a}|^2$ is removed.

\section{Numerical analysis of the effective potential with gluon condensates}

\subsection{Method}
 
We will perform the numerical analysis by following the approach 
developed in Ref.~\citen{Steiner:2002gx}.
The essence of this approach is the following.
In order to calculate the fermion determinant,
it is sufficient to know the energy eigenvalues of 
the inverse fermion propagator.
Then, after integrating over $p_0$ in the loop integral,
we obtain
\begin{equation}
 \int\frac{d^4 P}{i(2\pi)^4}\Tr\ln S_g^{-1}(P) =
 \frac{1}{2}\sum_{a=1}^n \int\frac{d^3 p}{(2\pi)^3}|E_a|,
\end{equation}
where $E_a$'s are the energy eigenvalues of a $n \times n$ matrix 
$S_g^{-1}(P)$ including both positive (particle) and negative 
(anti-particle) eigenvalues.

\begin{figure}[tb]
\parbox{\halftext}{\resizebox{0.47\textwidth}{!}{
\includegraphics{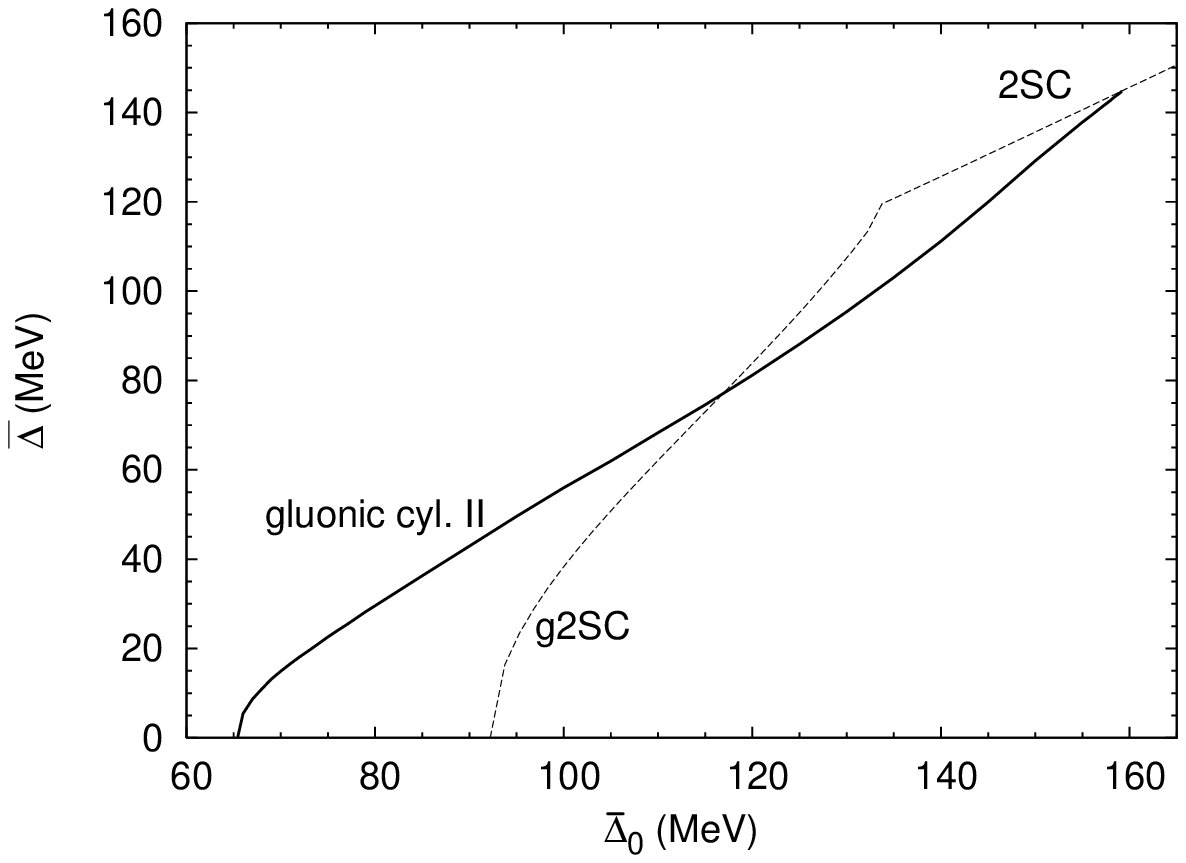}}
\caption{The diquark gap $\bar{\Delta}$ for the gluonic cylindrical phase II 
         and the 2SC/g2SC phases. At $\bar{\Delta}_0=134.6$ MeV, the g2SC 
         phase turns into the 2SC one.
         Here the values $\Lambda=653.3$ MeV and $\mu=400$ MeV were used.
\label{fig_d}}}
\hfil
\parbox{\halftext}{\resizebox{0.45\textwidth}{!}{
\includegraphics{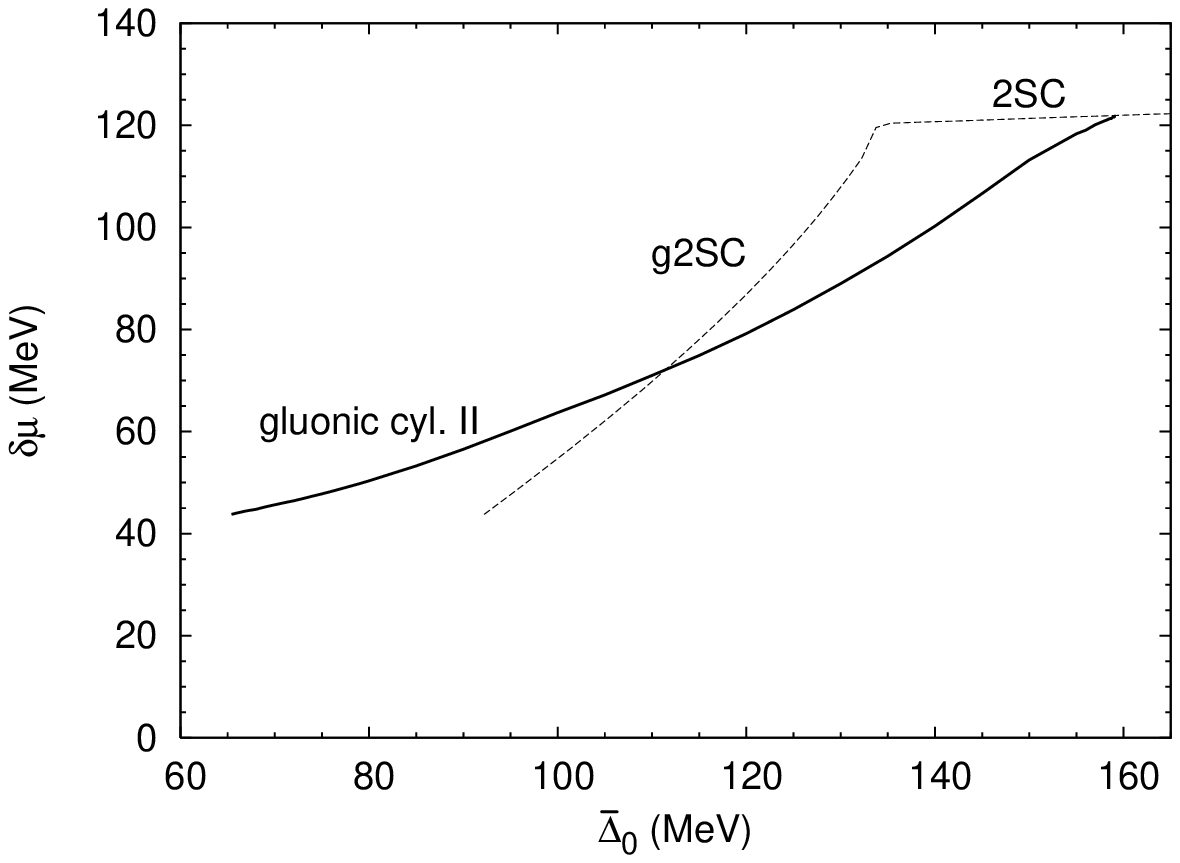}}
\caption{The mismatch $\delta\mu$ in the gluonic cylindrical phase II 
         and the 2SC/g2SC phases.
         At $\bar{\Delta}_0=134.6$ MeV, the g2SC phase turns into the 2SC one.
         The values $\Lambda=653.3$ MeV and $\mu=400$ MeV were used.
\label{fig_dm}}}
\end{figure}

Before explicit calculations, let us show how 
one can block-diagonalize 
the inverse propagator $S_g^{-1}$ 
in the flavor and chirality spaces: this will allow to reduce the
$48 \times 48$ $S_g^{-1}$  matrix to a $12 \times 12$ one.
We write: 
\begin{eqnarray}
  \det S_g^{-1} &=&
  \det \bigg[\,
  \left( \begin{array}{cc}\varepsilon & 0 \\ 0 & 1 \end{array}\right) 
  S_g^{-1} 
  \left( \begin{array}{cc} 1 & 0 \\ 0 & \varepsilon \end{array}\right) 
  \,\bigg] \\[3mm]
&=& \det \left(
  \begin{array}{cc}
   [G_{0,g}^+]^{-1} & +i\epsilon^b\gamma_5\Delta \\[2mm] 
  -i\epsilon^b\gamma_5\Delta &  [\tilde{G}_{0,g}^-]^{-1}
  \end{array}
  \right) ,
  \label{S_g_diag}
\end{eqnarray}
with
\begin{equation}
  [\tilde{G}_{0,g}^-]^{-1}(P) \equiv 
  (p_0-\tilde{\bm{\mu}}_0-\bm{\mu}_{\rm col}^T)\gamma^0
  - \vec p \cdot \vec \gamma + \vec A^{\;T} \cdot \vec \gamma,
\end{equation}
where we defined
\begin{equation}
  \bm{\mu}_{\rm col} \equiv g\VEV{A_0^a} T^a, \quad
  \vec A \equiv g\vec A^aT^a , \quad 
  \tilde{\bm{\mu}}_0 \equiv \bar{\mu}_0 + \delta\mu \tau_3 .
  \label{def-mu} 
\end{equation}
The expression (\ref{S_g_diag}) is flavor-diagonal. 

Now, since the current quark masses are neglected, the theory is 
chiral invariant.
Therefore one can decompose the inverse propagator into the 
right- and left-handed parts.
This reduces the determinant of the $48 \times 48$
matrix $S_g^{-1}$ to that of two $12 \times 12$ flavor-diagonal 
matrices. For the single plane wave LOFF phase, we can 
further reduce the size of the matrix to a $4 \times 4$ one.

\subsection{Results}

The form and the structure of the the effective potential (\ref{V_R}) 
were analyzed numerically.
We search for stationary points of the potential,
which correspond to the solutions of the neutrality conditions 
and the gap equations.
In the analysis, we take realistic values $\mu=400$ MeV and 
$\Lambda=653.3$ MeV.
The results are qualitatively unchanged in the region of
the parameters with $\mbox{300 MeV} < \mu < \mbox{500 MeV}$ and 
$\mbox{653.3 MeV} < \Lambda < \mbox{1 GeV}$.

\begin{wrapfigure}{l}{\halftext}
\centerline{\includegraphics[width=0.47\textwidth]{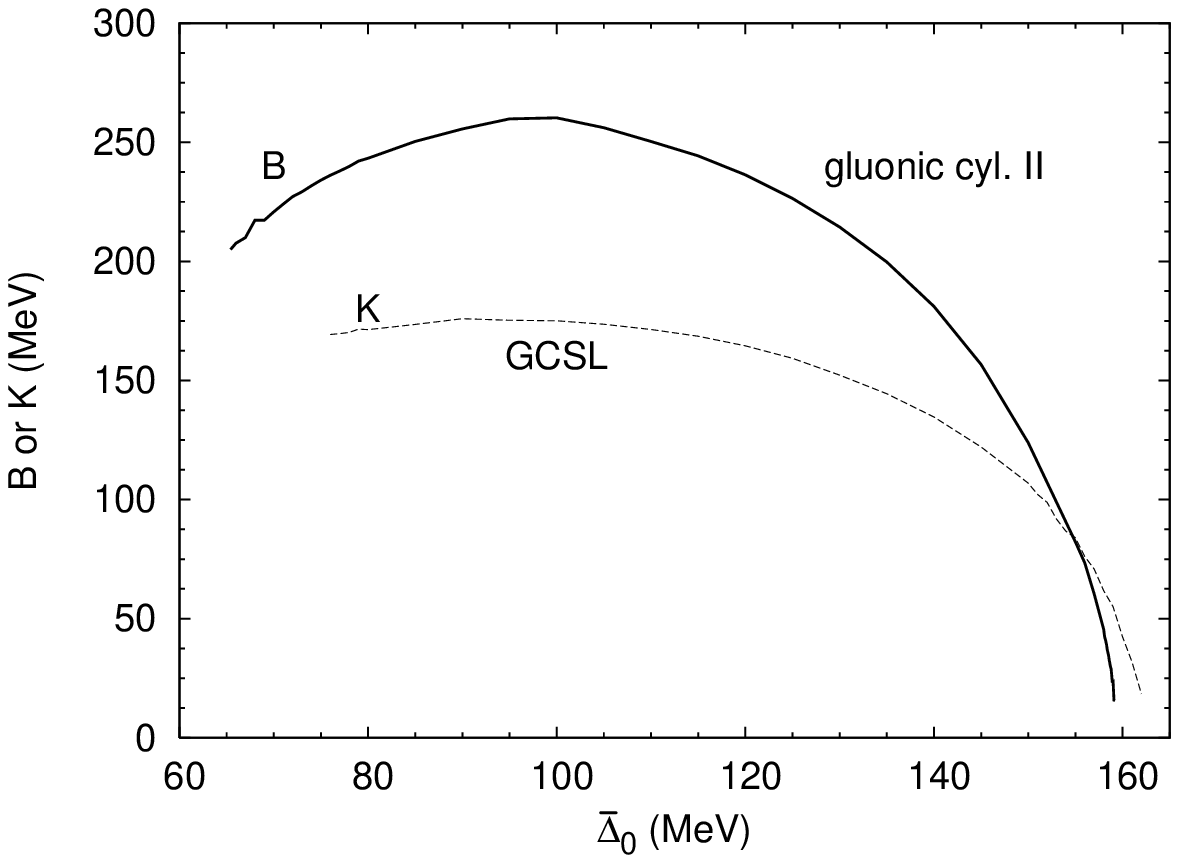}}
\caption{The gluon condensates $B$ and $K$ in
         the gluonic cylindrical phase II and the GCSL phase.
         The values $\Lambda=653.3$ MeV, $\mu=400$ MeV, and $\alpha_s=1$
         were used.}
\label{fig_b}
\end{wrapfigure}

We find that the gluonic cylindrical phase II does exist
in the region
\begin{equation}
 6.5 \times 10^1 \mbox{ MeV} < \bar{\Delta}_0 < 1.6 \times 10^2 \mbox{ MeV}\, . 
\end{equation}
In this case, the tree gluon term in Eq.~(\ref{V}) is negligible, so that 
the results are not sensitive to the choice of the value of the strong 
coupling $\alpha_s = g^2/4\pi$.
We depict the solutions for $\bar{\Delta}$ and $\delta\mu$, following
from the neutrality conditions,
in Figs.~\ref{fig_d} and \ref{fig_dm}, respectively. Note
that due to the neutrality conditions, 
the values of $\bar{\Delta}$ and $\delta\mu$
in the gluonic cylindrical phase II is significantly different from 
those in the 2SC/g2SC ones.
The behavior of the gluon condensate $B$ is presented in Fig.~\ref{fig_b}.
The phase transition from the gluonic cylindrical phase II to 
the 2SC one seems to be a weakly first order one. 
This conclusion is, however, not definite, because the accuracy is
not so good
near this critical point. On the other hand, there definitely exists 
a first order phase transition between the cylindrical phase and 
the normal one: as one can see in Fig.~\ref{fig_b}, the value
of the condensate $B$ is large at the starting point of the former. 

\begin{wrapfigure}{r}{\halftext}
\centerline{\resizebox{0.46\textwidth}{!}{
\includegraphics{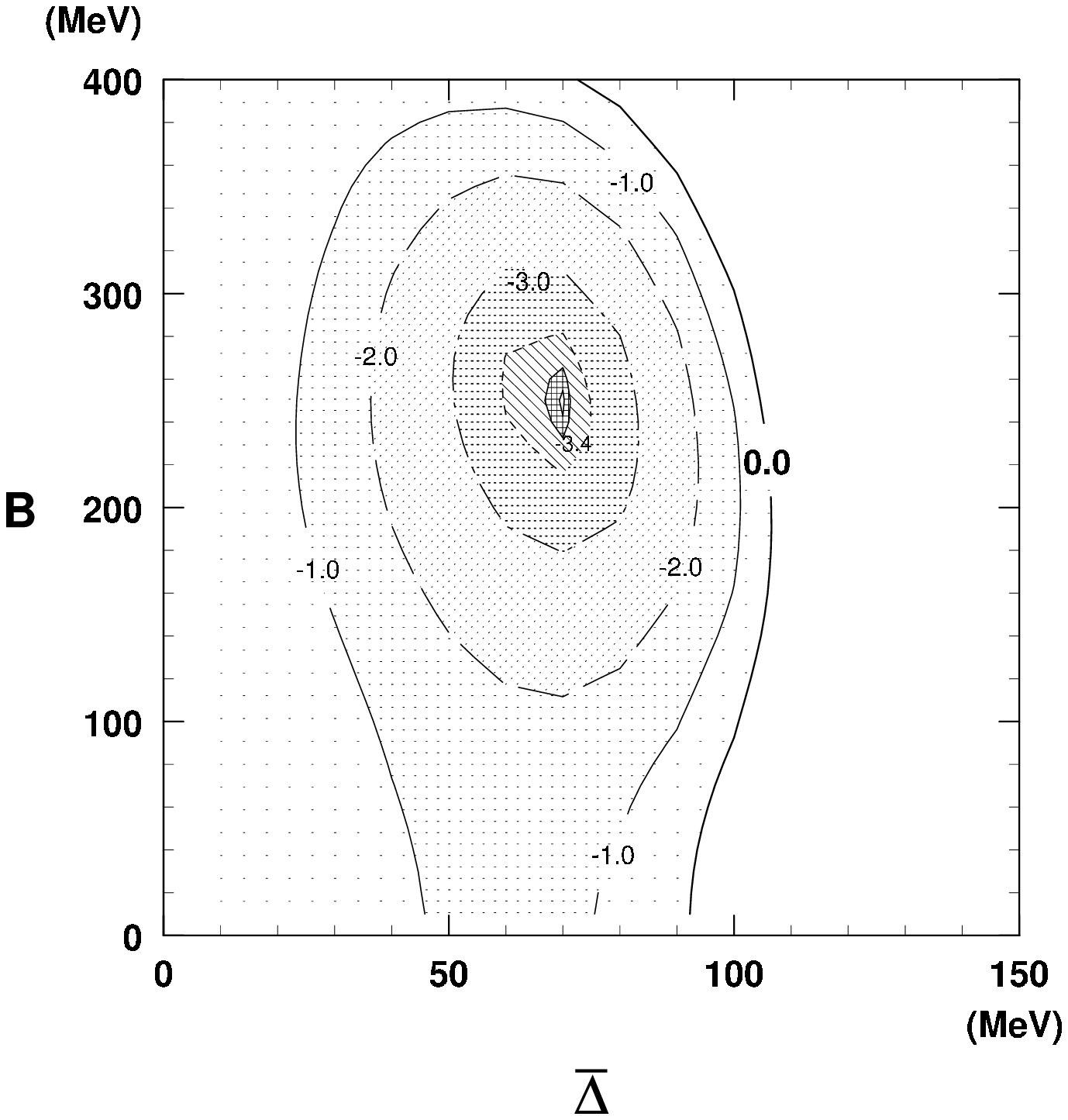}}}
\caption{The contour plot of the neutral effective potential 
          for the gluonic cylindrical phase II.
          Since the neutrality conditions for 
          $\delta\mu$, $\mu_3$ and $\mu_8$ were solved at each point 
          $(\bar{\Delta}, B)$,
          their values are different at different points.
          The reference point for the free energy is its value 
          in the normal phase. 
          The values $\bar{\Delta}_0=110$ MeV, $\Lambda=653.3$ MeV, 
          and $\mu=400$ MeV were used.}
\label{fig_con}
\end{wrapfigure}

In Fig.~\ref{fig_con}, 
the neutral effective potential at $\bar{\Delta}_0=110$ MeV is shown,
with $\delta\mu$, $\mu_3$ and $\mu_8$  
determined from the neutrality conditions.
It is clear that the solution for the 
gluonic cylindrical phase II corresponds to a minimum of 
the neutral effective potential.

The values of the color chemical potentials $\mu_3$ and $\mu_8$ is 
found to be at most $|\mu_3| \sim {\cal O}(\mbox{50 MeV})$ and 
$|\mu_8| \sim {\cal O}(\mbox{40 MeV})$ for 
$\mbox{300 MeV} < \mu < \mbox{500 MeV}$ and 
$\mbox{653.3 MeV} < \Lambda < \mbox{1 GeV}$.
At the starting point of this phase, the value of the gap
$\bar{\Delta}$ is zero.
At its endpoint,
the ratio of $\delta\mu/\bar{\Delta}$ is $\delta\mu/\bar{\Delta} = 
0.84\mbox{--}0.86 > 1/\sqrt{2}=0.707$
for these values $\mu$ and $\Lambda$.         
This result is consistent with that in Ref.~\citen{Kiriyama:2006xw}. 

Let us now turn to the results of the analysis for
the GCSL phase. It exists in the region
\begin{equation}
 7.6 \times 10^1 \mbox{ MeV} < \bar{\Delta}_0 < 1.6 \times 10^2 \mbox{ MeV} ,  
\end{equation}
for $\alpha_s=1$.
Note that the GCSL phase starts at a larger $\bar{\Delta}_0$ than
the gluonic cylindrical phase II.
The values of $\bar{\Delta}$ and $\delta\mu$ in these two phases
are not so different. However, unlike the case of the gluonic cylindrical 
phase II,
the gap $\bar{\Delta}$ at the starting point in the GCSL phase   
is quite big, which is a strong indication on a first order phase transition
between this phase and the normal one.

The form of the solution for 
the gluon condensate $K$ is depicted in Fig.~\ref{fig_b}.
Note that there is a tiny discrepancy ($\sim 0.5$\%) 
between the values of the endpoints of
these two gluonic phases.
A more accurate analysis is required to answer whether  
it is a real thing or an artifact. On the other hand, a large value
of the condensate $K$ at the starting point strongly supports
the conclusion made above that a first order phase 
transition takes place between the GCSL phase and the normal one.

\begin{wrapfigure}{l}{\halftext}
\centerline{\resizebox{0.47\textwidth}{!}{
\includegraphics{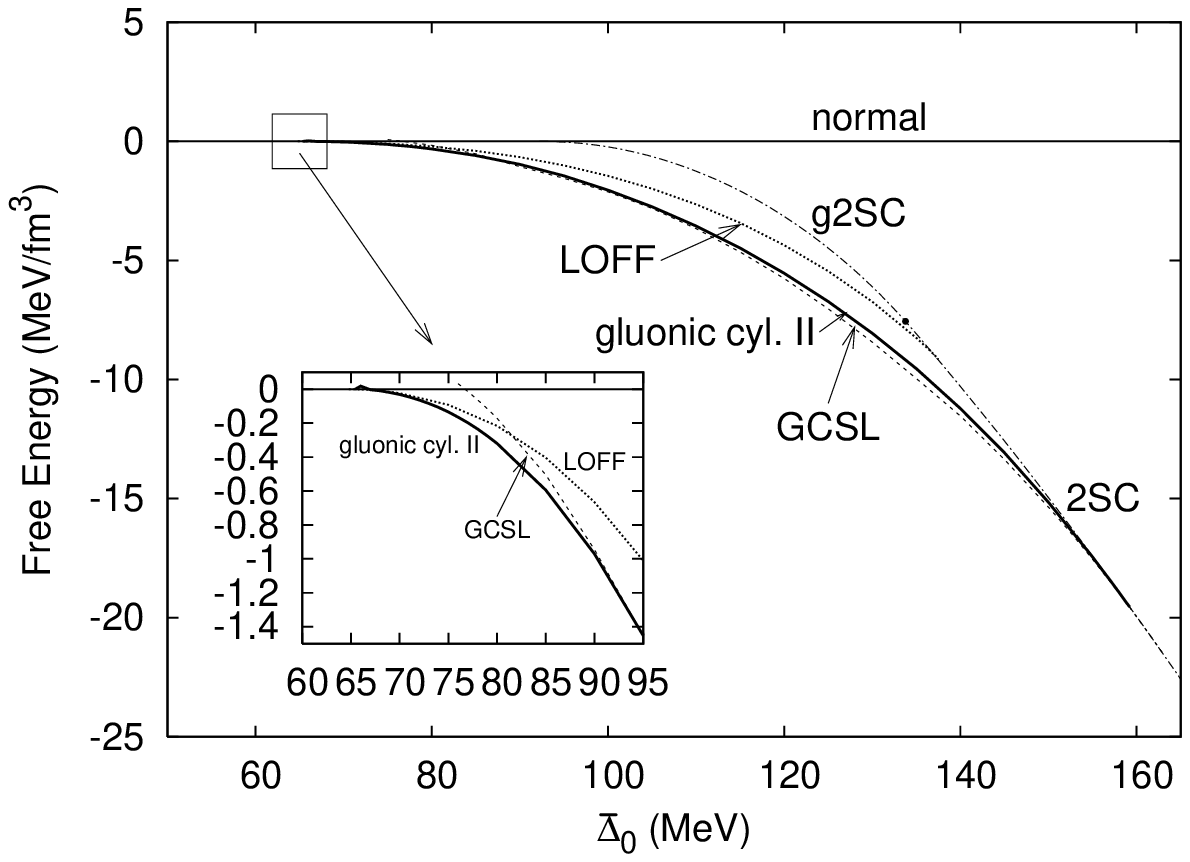}}}
\caption{The free energy in different phases. Its value in
          the normal phase is chosen as the reference point.
          The bold and dashed curves correspond to the gluonic 
          cylindrical and GCSL phases, respectively. 
          The dotted and dash-dotted ones correspond
          to the single plane wave LOFF phase and the 2SC/g2SC ones.
          The values $\Lambda=653.3$ MeV, $\mu=400$ MeV, and $\alpha_s=1$
          were used. In the inset, the free energy 
          in the region near the normal phase is shown.}
\label{fig2}
\end{wrapfigure}

Let us now compare the free energy for the 2SC/g2SC, LOFF, GCSL and
gluonic cylindrical phases.
The corresponding numerical results are shown in Fig.~\ref{fig2}.
In the weak coupling regime with  
$\bar{\Delta}_0 < 65\mbox{ MeV}$, the normal phase with no condensates
is realized.
The single plane wave LOFF phase is the most stable one in a
narrow window with $65\mbox{ MeV} < \bar{\Delta}_0 < 67\mbox{ MeV}$. 
The gluonic cylindrical/GCSL phases are energetically favorable
in a wide region with
$67\mbox{ MeV} < \bar{\Delta}_0 < 1.6 \times 10^2 \mbox{ MeV}$,
corresponding to the intermediate and strong coupling regimes.
At $\bar{\Delta}_0 > 1.6 \times 10^2 \mbox{ MeV}$ 
the 2SC phase is realized.

Comments
concerning the sensitivity of the dynamics in
the GCSL phase with respect to the strong coupling $\alpha_s$
are now in order.
In the region close to the endpoint
of this phase,
its free energy
$V_{\rm eff}$
is essentially independent of $\alpha_s$, i.e.,
the tree gluon term is irrelevant there.
On the other hand, the value of $\bar{\Delta}_0$ at
the starting point is rather sensitive to
the choice of $\alpha_s$.
For example,
\begin{equation}
  \bar{\Delta}_0^{\rm start}(\mbox{GCSL}) = 82, 76, 69 \mbox{ MeV} \;
\label{start}
\end{equation}
for $\alpha_s=0.75, 1.00, 1.25$.
The sensitivity of
the free energy $V_{\rm eff}$
to the choice of $\alpha_s$ is also higher
in that region, although it is not too strong:
the variations of $V_{\rm eff}$
are at most $0.4$ $\mbox{MeV}/\mbox{fm}^3$ for
$\alpha_s=0.75\mbox{--}1.25$.
The latter is comparable with the difference between
the free energies of the
LOFF and GCSL phases in the region $\bar{\Delta}_0 < 95\mbox{MeV}$
at $\alpha_s=1$ (see Fig.~\ref{fig2}).
The origin of the sensitivity in that region
is connected with a contribution of
the tree gluonic term in the effective potential (\ref{V}),
which is provided
by a chromomagnetic field strength condensate.
At the starting point,
it is
$g\VEV{F_{yz}} = K^2/2 \sim (\mbox{110--130 MeV})^2$ for
$\alpha_s=0.75\mbox{--}1.25$.
It is important
that, despite this sensitivity,
qualitative results of the analysis in this region,
such as a strong first order phase transition at the starting point,
etc., are robust.

As one can see in Fig.~\ref{fig2},
there exists a critical value $\bar{\Delta}_{0}^{\rm cr}$ 
in the region 
$76\mbox{ MeV} < \bar{\Delta}_0 < 1.6 \times 10^2 \mbox{ MeV}$
dividing the gluonic cylindrical and GCSL phases: the latter
becomes more stable at $\bar{\Delta}_0 > \bar{\Delta}_{0}^{\rm cr}$.
However, since the value of $\bar{\Delta}_{0}^{\rm cr}$ is sensitive
to the choice of $\alpha_s$, we do not give here a concrete number
for it.

It is instructive to compare our results with those in 
Ref.~\citen{Kiriyama:2006ui}, where
(in the present language) the gluonic
cylindrical phase II was also numerically analyzed, but by
utilizing the approximations described in the Introduction.
In both these analyses, this gluonic phase is energetically more
favored than the single plane wave LOFF one
in a wide range of the parameter space. On the other hand,
unlike~\citen{Kiriyama:2006ui}, the gluonic phase
in our analysis appears essentially earlier than
the g2SC phase which exists only in the region
$92.2\mbox{ MeV} < \bar{\Delta}_0 < 134.6\mbox{ MeV}$.
This feature is qualitatively unchanged for
$300\mbox{ MeV} < \mu < 500\mbox{ MeV}$ and
$653.3\mbox{ MeV} < \Lambda < 1\mbox{ GeV}$.
It strongly suggests that the gluonic phase can resolve
the problem of the 2SC/g2SC chromomagnetic instability
in the whole region where it exists.

\section{Summary and discussions}

The main result of this paper is establishing the existence
of the gluonic cylindrical phase II and the GCSL one 
that led us to the suggestion of the form of the
phase diagram in the cold two flavor dense quark matter.
These phases are energetically more 
favorable than the 2SC/g2SC and single plane wave LOFF ones 
in a wide region of the parameter space.
This indicates that the gluonic phases can resolve 
the problem of 
the chromomagnetic and plasmon instabilities in two-flavor
dense QCD.
We also found that the single plane wave LOFF state is
energetically favorable in a window
$65\mbox{ MeV} < \bar{\Delta}_0 < 67\mbox{ MeV}$.
In this region, the LOFF state is probably free from
the chromomagnetic instability.\cite{Gorbar:2005tx}
It would be also worthwhile to examine the multiple plane wave LOFF
state.\cite{Bowers:2002xr} The 2SC state and the
normal one are realized in the strong and weak coupling
regimes, respectively, as was expected.

There is still quite a number of questions that need to be answered.
It is still a nontrivial problem whether or not the $v^2 < 0$ instability 
in the diquark Higgs mode\cite{Hashimoto:2006mn,Iida:2006df,Giannakis:2006gg} 
is absent in the gluonic phases. 
It would be also important to calculate explicitly the Meissner masses
squared in order to prove
the absence of the chromomagnetic instability there.
Note that their values
are essentially {\it different} from the eigenvalues of
the curvature of the neutral effective potential.
Indeed, while the former are calculated at fixed values of
the chemical potentials $\delta\mu$, $\mu_{3}$ and $\mu_8$ (corresponding
to the given gap $\bar{\Delta}$ and the condensates $B$ or $K$), the latter
are calculated by varying the effective potential
when these chemical potentials are also varying with $\bar{\Delta}$
and $B$ (or $K$) through the neutral conditions. 
This problem will be considered elsewhere.

The dynamics with vector condensates of gauge fields, such as
in the gluonic phases, are very rich and
one can expect intriguing phenomena there, such as the existence of 
exotic hadrons,\cite{Gorbar:2005rx,Gorbar:2007vx} \ 
roton-like excitations,\cite{sigmamodel} \
and vortex-like ones.\cite{Gorbar:2005pi} The GCSL phase is
especially interesting: it describes an anisotropic
superconductor. Moreover, we expect an 
abnormal number of gapless Nambu-Goldstone (NG) bosons
in its spectrum. This expectation is based on recent
results in Ref.~\citen{Buchel:2007rx} obtained in the gauged $\sigma$-model
with chemical potential for hypercharge. In that work,
a phase with vector
condensates, whose properties are very similar to those in the 
GCSL one, has been revealed (it was called the gauge-flavor-spin locked phase).
It was shown that there is an abnormal number of NG bosons in that phase.
\footnote{It is the same phenomenon as that found in a non-gauge
relativistic model at finite density in Ref.~\citen{Miransky:2001tw}.}
We hope to turn to these dynamical issues elsewhere.

\section*{Acknowledgments}

M.H. acknowledges useful discussions with H. Abuki.
A part of this work was done when V.A.M. visited Yukawa Institute
for Theoretical Physics, Kyoto University. He is grateful to Professor
Taichiro Kugo and Professor Teiji Kunihiro for their warm hospitality.
The research of M.H. was supported by the JSPS Grant-in-Aid for
Scientific Research (B) 18340059.
The work of V.A.M. was supported by the Natural Sciences and Engineering
Research Council of Canada.

\end{document}